\documentclass{article}
\usepackage{spconf,amsmath,amssymb, graphicx,hyperref}

\title{Omni-directional attention mechanism based on Mamba for speech separation}
\name{Ke~Xue$^{1}$, Chang~Sun$^{2}$, Rongfei~Fan$^{1,\dagger}$, Jing~Wang$^{3}$, Han~Hu$^{3}$\thanks{$^{\dagger}$ Corresponding author.}}
\address{$^{1}$School of Cyberspace Science and Technology, Beijing Institute of Technology, Beijing 100081, China\\
$^{2}$School of Computer Science, Beijing University of Posts and Telecommunications, Beijing 100876, China\\
$^{3}$School of Information and Electronics, Beijing Institute of Technology, Beijing 100081, China\\
\texttt{xueke924@bit.edu.cn,sunch@bupt.edu.cn} \\
\texttt{\{fanrongfei, wangjing, hhu\}@bit.edu.cn}}

\begin{document}
%
\maketitle
\begin{abstract}
Mamba, a selective state-space model (SSM), has emerged as an efficient alternative to Transformers for speech modeling, enabling long-sequence processing with linear complexity. While effective in speech separation, existing approaches, whether in the time or time-frequency domain, typically decompose the input along a single dimension into short one-dimensional sequences before processing them with Mamba, which restricts it to local 1D modeling and limits its ability to capture global dependencies across the 2D spectrogram. In this work, we propose an efficient omni-directional attention (OA) mechanism built upon unidirectional Mamba, which models global dependencies from ten different directions on the spectrogram. We expand the proposed mechanism into two baseline separation models and evaluate on three public datasets. Experimental results show that our approach consistently achieves significant performance gains over the baselines while preserving linear complexity, outperforming existing state-of-the-art (SOTA) systems.
\end{abstract}


\section{INTRODUCTION}
\label{sec:intro}

Speech separation (SS) aims to extract clean speech signals of individual speakers from a mixture with noise and reverberation. In recent years, with the rapid development of machine learning, deep learning-based speech separation methods \cite{wang2018supervised} have achieved remarkable progress. Existing approaches can be broadly categorized into three types: convolutional neural networks (CNN) \cite{lecun1989backpropagation}, recurrent neural networks (RNN) \cite{hochreiter1997long}, and transformer-based methods \cite{vaswani2017attention}, which was originally proposed for sequence-to-sequence modeling.



\begin{figure*}[htb]
\begin{minipage}[b]{ \linewidth}
 \centering
 \centerline{\includegraphics[width= 0.8 \textwidth]{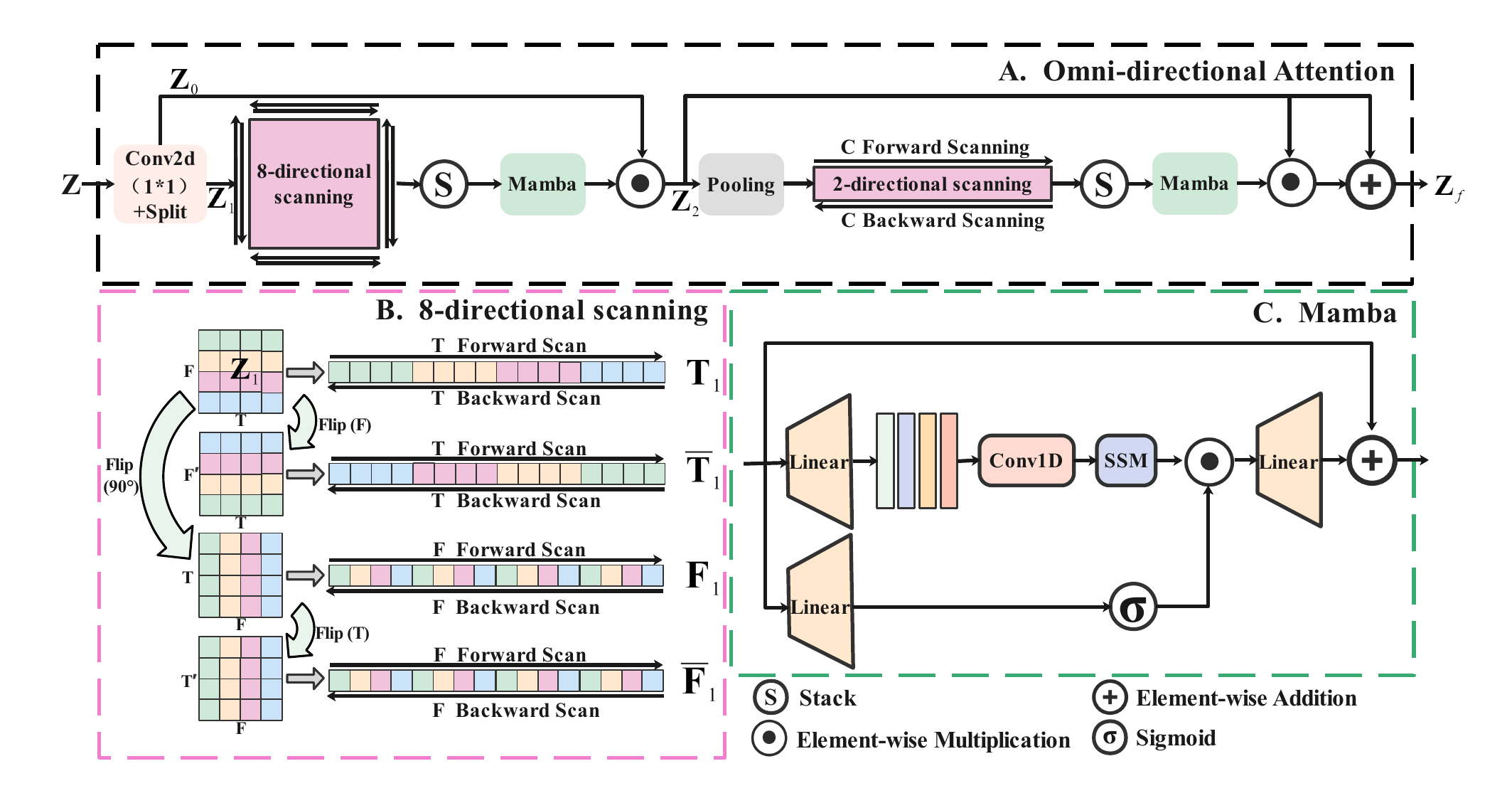}}
\end{minipage}

\vspace{-0.5cm}
\caption{The block diagram of the A. Omni-directional attention, B. 8-directional scanning, and C. Mamba} 
\label{f:fig1}
\end{figure*}

Early works introduced CNNs \cite{luo2019conv, tzinis2020sudo, li2022efficient, hu2021speech} into speech separation for their efficiency and low latency, but their limited receptive fields restrict long-range modeling. RNNs \cite{luo2020dual, li2022use} were then adopted to capture temporal dependencies, with the dual-path recurrent neural network (DPRNN) \cite{luo2020dual} achieving strong performance through intra- and inter-chunk modeling. However, RNNs remain inefficient and poorly parallelizable. Transformers \cite{chen2020dual, subakan2021attention, yang2022tfpsnet} further advanced separation by leveraging self-attention to model global dependencies, significantly outperforming CNN- and RNN-based systems. Yet, their quadratic complexity with sequence length makes them costly for long speech. Thus, developing models that balance long-range modeling ability with computational efficiency remains a key challenge.

State space models (SSMs) \cite{kalman1960new, gu2021combining, gu2021efficiently, gu2022parameterization} have emerged as an effective approach to address the aforementioned challenge, which has linear complexity, taking 1D sequences as input and leveraging hidden state recursion to capture long-range dependencies. Representative models such as S4 \cite{gu2021efficiently} and S5 \cite{gu2022parameterization} laid the foundation for this line of work. Building upon them, Mamba \cite{gu2023mamba}, a type of SSM, introduces a selective scan mechanism that further improves representational capacity and efficiency while preserving linear complexity. Mamba has demonstrated performance surpassing Transformers across various domains, including natural language processing \cite{pioro2024moe, yang2024clinicalmamba}, visual modeling \cite{liu2024vmamba, zou2024wave, shi2025vmambair}, and speech processing \cite{miyazaki2024exploring, wang2025mamba}.

Motivated by these advantages, recent studies \cite{chen2023neural, jiang2025dual, dang2024u, avenstrup2025sepmamba, li2024spmamba} have applied Mamba to speech separation. DPMamba \cite{jiang2025dual} integrates bidirectional Mamba into the dual-path framework, replacing Transformers to reduce parameters and computational cost while maintaining performance. U-Mamba-Net \cite{dang2024u} and SepMamba \cite{avenstrup2025sepmamba} adopt lightweight U-shaped Mamba architectures, showing robust separation under noise and reverberation. SPMamba \cite{li2024spmamba}, built on the TF-GridNet \cite{wang2023tf}, replaces BLSTMs with bidirectional Mamba modules, achieving fewer parameters and improved performance, further highlighting Mamba’s effectiveness.


Existing approaches, whether in the time or time-frequency domain, typically split the input along one dimension into short one-dimensional sequences before processing with Mamba. This local processing may prevent Mamba from directly modeling global structures across the full two-dimensional spectrogram. To address this, we propose an efficient omni-directional attention (OA) mechanism built upon Mamba.
The contributions of this work are threefold:

\begin{itemize}

\item We propose an omni-directional mapping that projects the spectrogram into 10 one-dimensional directions across the temporal (T), frequency (F), and channel (C) dimensions, enabling comprehensive global feature representation.  

\item To fully leverage these 10 1D directions derived from the 2D time-frequency spectrum, we design an OA mechanism that employs Mamba to model long-range dependencies across the entire spectrogram.


\item We build auxiliary architectural components around OA, integrating it into TF-GridNet and SPMamba to fully exploit its potential.
\end{itemize}

Experimental results on WSJ0-2Mix, WHAM!, and Libri2Mix demonstrate that our approach achieves performance gains, reaching state-of-the-art (SOTA) levels while maintaining linear computational complexity, validating the effectiveness of the omni-directional attention mechanism.

\section{METHOD}
\label{sec:format}

In this section, we first introduce the backbone of our attention mechanism, namely the state space model, Mamba. We then present the proposed omni-directional attention mechanism in detail which is our core module. Finally, we describe how this mechanism is integrated into SS models, TF-GridNet and SPMamba.

\subsection{Selective state space model (Mamba)}
\label{sec:format}

Mamba is a selective state space model (selective SSM) implementation, inspired by traditional state space model (SSM), as shown in Fig.\ref{f:fig1}-C.
It accumulates information through hidden states across the sequence, enabling conditional and selective propagation or forgetting of information for the current input. Mamba is accompanied by a parallel scan algorithm optimized for modern GPUs, which allows content-sensitive, linear-time modeling of long sequences. 

In a continuous-time system, the input $x(t)$ is processed through a high-dimensional state space $h(t) \in \mathbb{R}^H$ across all time steps $T$, which can be formulated as:

\begin{equation}
h'(t) = A h(t) + B x(t), \quad y(t) = C^\top h(t),
\end{equation}

where $A \in \mathbb{R}^{H \times H}$ is the state transition matrix, $B \in \mathbb{R}^{H \times 1}$ is the input matrix, $C \in \mathbb{R}^{1 \times H}$ is the output projection matrix, and $H$ denotes the hidden state dimension. To implement this system on digital computers, the continuous system is discretized as:
\begin{equation}
h_t = \bar{A} h_{t-1} + \bar{B} x_t, \quad y_t = C^\top h_t.
\end{equation}
where $\bar{A}$ and $\bar{B}$ are approximated using a zero-order hold:
\begin{equation}
\bar{A} = \exp(\Delta A), \quad \bar{B} = (\Delta A)^{-1} (\exp(\Delta A) - I) \cdot \Delta B.
\end{equation}
Under this discretization, the output sequence $y$ can be viewed as the input sequence $x_t$ convolved with a structured kernel $K$ composed of $A$, $B$, and $C$:
\begin{equation}
\bar{K} = \big(C \,\bar{B},\, C \,\overline{A B},\, \dots,\, C \,\overline{A^{L-1} B}\big), \quad y = x \ast \bar{K}.
\end{equation}
Reference~\cite{gu2023mamba} proposes an efficient hardware-friendly scan algorithm that significantly accelerates this computation.

\subsection{Omni-directional attention (OA) mechanism}
\label{sec:format}

We propose an OA mechanism built upon the unidirectional Mamba module to capture long-range dependencies in two-dimensional time-frequency representations, as illustrated in Fig.~\ref{f:fig1}-A. Given an input tensor $Z \in \mathbb{R}^{B \times C \times F \times T}$, where $B$, $C$, $F$, and $T$ denote the batch size, channel number, frequency bins, and time steps, respectively, we first apply a $1 \times 1$ 2D convolution to double the channel dimension and split it evenly into two parts. The second part, $Z_1$, is processed by an 8-directional scanning module, shown in Fig.~\ref{f:fig1}-B, which performs forward and backward scanning along both time and frequency dimensions. The resulting directional features are stacked and fed into a Mamba module. After summation along the directional dimension, the output is fused with $Z_0$ via element-wise multiplication:
\begin{align}
    [Z_0, Z_1] &= \text{Conv}_{1\times1}(Z), \\
    U &= \text{Stack}(T_1, \bar{T_1}, F_1, \bar{F_1}), \\
    Z_2 &= Z_0 \odot \text{Sum}(\text{Mamba}(U)).
\end{align}

A 2D pooling operation is then applied to $Z_2$ to obtain a compact representation:
\begin{equation}
    Z_2' = \text{Pooling}(Z_2) \in \mathbb{R}^{B \times 1 \times C}.
\end{equation}

2-directional scanning is performed along the channel dimension:
\begin{equation}
    C_1 = \text{Scan}_C(Z_2'), \quad 
    \bar{C_1} = \text{Scan}_C(\text{Flip}_C(Z_2')).
\end{equation}

The two directional features are stacked and processed by another Mamba module:
\begin{equation}
    Z_2'' = \text{Mamba}(\text{Stack}(C_1, \bar{C_1})).
\end{equation}

Finally, the channel-wise output is combined with $Z_2$ through element-wise multiplication and a residual connection to form the final representation:
\begin{equation}
    Z_f = Z_2 \odot Z_2'' + Z_2.
\end{equation}

\subsection{Incorporation into SS models}
\label{sec:format}

We integrate the proposed OA mechanism into TF-GridNet \cite{wang2023tf} and SPMamba \cite{li2024spmamba}, both of which consist of a frequency-domain module, a time-domain module, and a full-band self-attention module. Specifically, we replace the original self-attention with OA and explore three integration strategies: 1) applying OA at the front-end of the separation module; 2) at the back-end; 3) both front-end and back-end simultaneously. The effectiveness of these strategies is evaluated in the ablation study, with the third strategy used as the default in all standard experiments.

\section{EXPERIMENTAL SETUP}
\label{sec:pagestyle}

To maintain consistency with previous studies, we conducted training and evaluation on three publicly available datasets: WSJ0-2Mix \cite{hershey2016deep}, WHAM! \cite{wichern2019wham}, and Libri2Mix \cite{cosentino2020librimix}. WSJ0-2Mix is derived from the WSJ0 corpus and includes 30 hours of training data (20k utterances), 8 hours of validation data (5k utterances), and 5 hours of test data (3k utterances). WHAM! consists of simulated mixtures of clean speech from the WSJ0 corpus, combined with various types of background noise at different signal-to-noise ratios. Libri2Mix is randomly sampled from the train-100 subset of the LibriSpeech dataset with an 8 kHz sampling rate. Each mixture is uniformly sampled in Loudness Units relative to Full Scale (LUFS) between -25 and -33 dB. The training set contains 13.9k utterances totaling 43 hours, while the validation and test sets each contain 3k utterances totaling 4 hours. All datasets are maintained at an 8 kHz sampling rate.

In the OA mechanism, the feature dimension is set to $C = 24$, and the hidden dimension of Mamba is 16 in TF-GridNet \cite{wang2023tf},while in SPMamba, $C = 48$, the hidden dimension is 48. All other configurations follow the parameter settings of TF-GridNet \cite{wang2023tf}. During each training epoch, a 4-second segment is randomly sampled from each mixture for model training. The optimizer is AdamW, and gradient clipping is applied with a norm of 5. The initial learning rate is 0.001 and is halved if the validation loss does not improve within 3 epochs. The evaluation metrics are the improvement of scale-invariant signal-to-noise ratio (SI-SDRi) \cite{le2019sdr} and the improvement of signal-to-distortion ratio (SDRi), consistent with previous studies. The number of model parameters and MACs were calculated using \texttt{ptflops}. 

\section{EVALUATION RESULTS}
\label{sec:results}

\subsection{Comparison with SOTA methods}
\label{SOTA methods}

We evaluated the effect of incorporating the Omni-Attention (OA) mechanism into two types of base models on the WSJ0-2Mix dataset and compared the results with other SOTA models, as shown in Table~\ref{tab:sota}. The SOTA models are divided into two groups: models without Mamba and models with Mamba. It is evident that models using Mamba generally outperform those without it. Among them, TF-GridNet and SPMamba achieve the best performance in their respective categories. After integrating our proposed OA mechanism, both TF-GridNet and SPMamba show further improvements, reaching the current SOTA level and demonstrating the effectiveness of the OA mechanism.

\begin{table}[h]
\centering
\caption{Comparison methods on WSJ0-2Mix. $^{*}$ indicates the 8.0M version of TF-GridNet is used, as the 14.4M version yields no significant performance gain.}
\label{tab:sota}
\resizebox{\linewidth}{!}{
\begin{tabular}{lccc}
\hline
\textbf{Methods} & \textbf{Params (M)} & \textbf{SI-SDRi (dB)} & \textbf{SDRi (dB)} \\
\hline
Conv-TasNet \cite{luo2019conv} & 5.1 & 15.3 & 15.6 \\ 
SudoRM-RF \cite{tzinis2020sudo} & 2.7 & 18.9 & 19.1 \\ 
TDANet \cite{li2022efficient} & 2.3 & 18.5 & 18.7 \\
A-FRCNN \cite{hu2021speech} & 6.1 & 18.3 & 18.6 \\ 
DPRNN \cite{luo2020dual} & 2.6 & 18.8 & 19.0 \\ 
DPTNet \cite{chen2020dual} & 2.7 & 20.2 & 20.6 \\ 
SepFormer \cite{subakan2021attention} & 26.0 & 20.4 & 20.5 \\
TFPSNet \cite{yang2022tfpsnet} & 2.7 & 21.1 & 21.3 \\
TF-Gridnet \cite{wang2023tf}$^{*}$ & 8.0 & 22.9 & - \\
\hline
S4M \cite{chen2023neural} & 3.6 & 20.5 & 20.7 \\ 
DPMamba \cite{jiang2025dual} & 15.9 & 22.6 & 22.7 \\ 
SepMamba \cite{avenstrup2025sepmamba} & 7.2 & 21.2 & 21.4 \\ 
SPMamba \cite{li2024spmamba} & 6.1 & 22.7 & 22.8 \\
\hline
\textbf{TF-Gridnet (OA)$^{*}$} & \textbf{8.1} & \textbf{23.1} & \textbf{23.3} \\ 
\textbf{SPMamba (OA)} & \textbf{7.2} & \textbf{23.0} & \textbf{23.1} \\
\hline
\end{tabular}
}
\end{table}

Table~\ref{tab:sota2} presents the performance of the models on more challenging datasets. Similarly, we apply the Omni-Attention (OA) mechanism to both types of base models. As shown, after incorporating OA, TF-GridNet achieves SI-SDRi improvements of 0.2 dB and 0.3 dB on the WHAM! and Libri2Mix datasets, respectively. Likewise, SPMamba gains improvements of 0.3 dB and 0.1 dB on the two datasets. These results further validate the effectiveness of the proposed OA mechanism.

\begin{table}[h]
\centering
\caption{Comparison methods on WHAM! and Libri2Mix}
\label{tab:sota2}
\resizebox{\linewidth}{!}{
\begin{tabular}{ccccc}
\hline
\textbf{Methods} & \multicolumn{2}{c}{\textbf{WHAM!}} & \multicolumn{2}{c}{\textbf{Libri2Mix}} \\ \cline{2-5}
                 & \textbf{SI-SDRi} & \textbf{SDRi} & \textbf{SI-SDRi} & \textbf{SDRi} \\ \hline
Conv-TasNet \cite{luo2019conv} & 12.7 & 13.0 & 12.2 & 12.7 \\
SudoRM-RF \cite{tzinis2020sudo} & 13.7 & 14.1 & 14.0 & 14.4 \\
TDANet \cite{li2022efficient} & 15.2 & 15.4 & 17.4 & 17.9 \\
A-FRCNN \cite{hu2021speech} & 14.5 & 14.8 & 16.7 & 17.2 \\
DualPathRNN \cite{luo2020dual} & 13.7 & 14.1 & 16.1 & 16.6 \\ \hline
TF-GridNet(14.4M) & 16.9 & 17.2 & 19.8 & 20.1 \\
SPMamba & \textbf{17.4} & \textbf{17.6} & {19.9} & {20.4} \\ \hline

\textbf{TF-GridNet(OA)} & {17.1} & {17.2} & \textbf{20.1} & \textbf{20.2} \\
\textbf{SPMamba(OA)} & \textbf{17.7} & \textbf{17.9} & \textbf{20.0} & \textbf{20.2} \\ \hline
\end{tabular}}
\end{table}

\subsection{Complexity comparison}
\label{Complexity}

\begin{figure}[htb]
\begin{minipage}[b]{1.0 \linewidth}
 \centering
 \centerline{\includegraphics[width=0.9 \columnwidth]{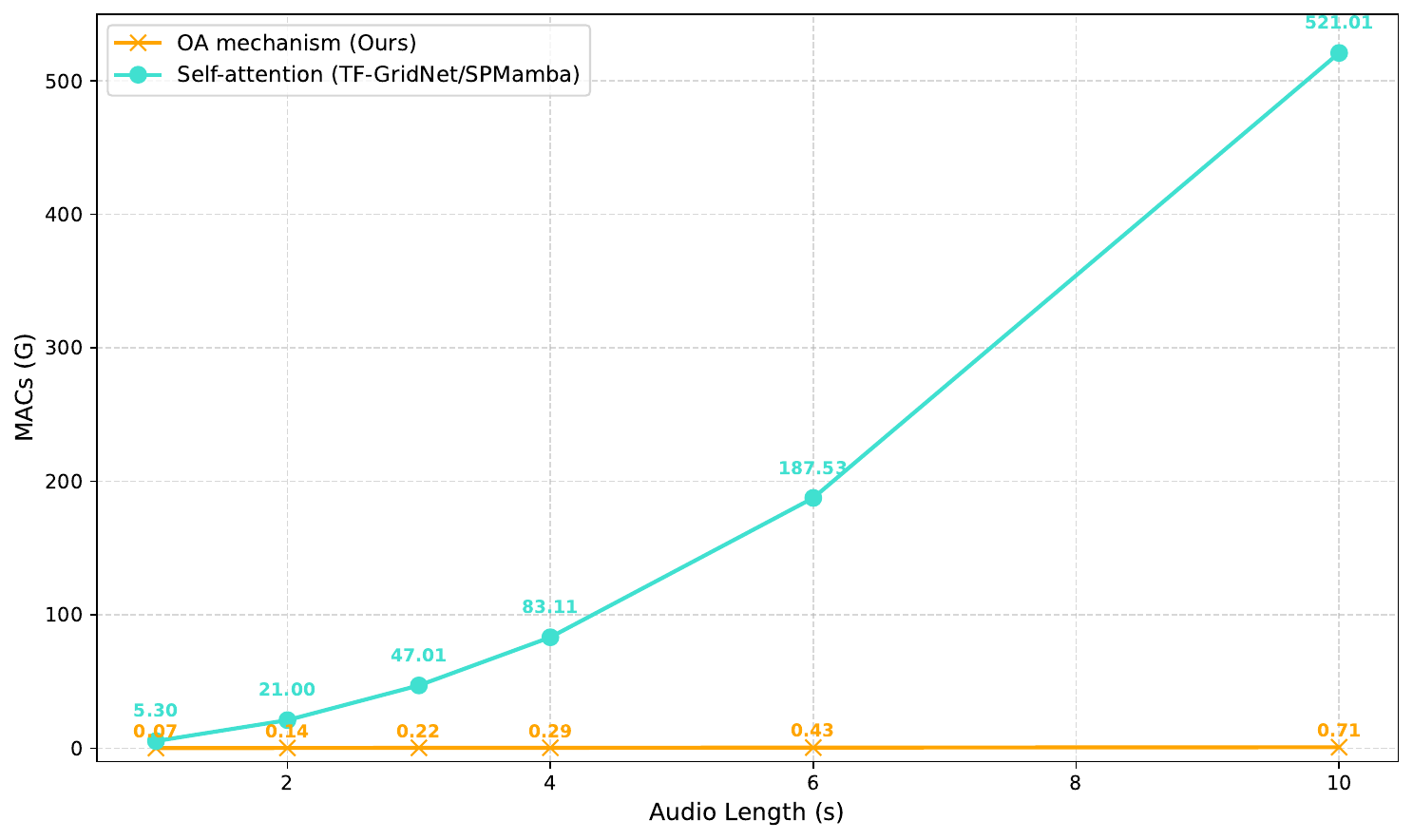}}
\end{minipage}

\vspace{-0.5cm}
\caption{Complexity comparisons between OA mechanism and self-attention in TF-GridNet and SPMamba.}
\label{f:fig0}
\end{figure}

\noindent
Figure~\ref{f:fig0} compares the computational complexity between the proposed OA mechanism and the self-attention mechanisms in two baseline models. As shown, the Multiply-Accumulate Operations (MACs) of the self-attention mechanisms grow exponentially with increasing audio length, significantly exceeding those of our proposed OA mechanism. Specifically, the two baseline models operate at the frame level, and the memory cost of their attention matrices is 
\(\mathcal{O}(B \times L \times T^{2})\), 
which is quadratic with respect to time. In contrast, the computational complexity of our OA mechanism remains at 
\(\mathcal{O}(B \times L \times T \times F)\), 
scaling only linearly with time, thereby demonstrating substantial efficiency advantages especially for long sequences.

\subsection{Ablation studies}
\label{Ablation studies}

Table~\ref{tab:ablation} illustrates the impact of different OA configurations on the model performance. By comparing various numbers of scanning directions, we observe that C scanning has a significant effect on performance. Specifically, when the number of Temporal and Frequency (T+F) scanning directions is kept the same, introducing C scanning leads to an improvement of about 0.2 dB in SI-SDR. In contrast, increasing the number of T+F scanning directions from 4 to 8 yields only a marginal gain of 0.1 dB. Moreover, applying the OA mechanism to both the front-end and back-end provides additional improvements compared to applying it to either the front-end or the back-end alone.

\begin{table}[t]
\centering
\caption{Ablation study on different configurations on WSJ0-2Mix using TF-Gridnet(OA).}
\begin{tabular}{lccc}
\hline
\textbf{Configuration} & \textbf{SI-SDRi (dB)} & \textbf{SDRi (dB)} \\
\hline
4-directional(T+F)          & 22.7 & 22.9 \\
6-directional(T+F+C)          & 23.0 & 23.2  \\
8-directional(T+F)            & 22.8 & 22.9  \\
\textbf{10-directional(T+F+C)}  & \textbf{23.1} & \textbf{23.3}  \\ \hline

front-end(OA)              & 23.0 & 23.1  \\
back-end(OA)              & 23.0 & 23.2  \\
\textbf{both(Default)} & \textbf{23.1} & \textbf{23.3}  \\
\hline
\end{tabular}
\label{tab:ablation}
\end{table}

\vspace{-0.3cm}

\section{CONCLUSION}
\label{sec:conclusion}

In this work, we proposed a linear-complexity Omni-directional Attention (OA) mechanism, which captures global features from the temporal (T), frequency (F), and channel (C) dimensions across ten directions. By integrating the OA mechanism into two baseline models, we achieved significant performance improvements. 




 




\vspace{-0.3cm}
\section{ACKNOWLEDGMENT}
\label{sec:acknowledgment}

This work was supported by the National Key Research and Development Program of China, the National Natural Science Foundation of China (No. 62571037), and the Beijing Natural Science Foundation (Nos. L257001 and L242089).

\vfill\pagebreak




\begin{footnotesize}
\bibliographystyle{IEEEtran}
\bibliography{strings,refs}

@inproceedings{tzinis2020sudo,
  title={Sudo rm-rf: Efficient networks for universal audio source separation},
  author={Tzinis, E. and Wang, Z. and Smaragdis, P.},
  booktitle={2020 IEEE 30th International Workshop on Machine Learning for Signal Processing (MLSP)},
  pages={1--6},
  year={2020},
  organization={IEEE}
}

@inproceedings{luo2020dual,
  title={Dual-path rnn: efficient long sequence modeling for time-domain single-channel speech separation},
  author={Luo, Y. and Chen, Z. and Yoshioka, T.},
  booktitle={ICASSP 2020-2020 IEEE International Conference on Acoustics, Speech and Signal Processing (ICASSP)},
  pages={46--50},
  year={2020},
  organization={IEEE}
}

@inproceedings{subakan2021attention,
  title={Attention is all you need in speech separation},
  author={Subakan, C. and Ravanelli, M. and Cornell, S. and Bronzi, M. and Zhong, J.},
  booktitle={ICASSP 2021-2021 IEEE International Conference on Acoustics, Speech and Signal Processing (ICASSP)},
  pages={21--25},
  year={2021},
  organization={IEEE}
}

@inproceedings{yang2022tfpsnet,
  title={TFPSNet: Time-frequency domain path scanning network for speech separation},
  author={Yang, L. and Liu, W. and Wang, W.},
  booktitle={ICASSP 2022-2022 IEEE International Conference on Acoustics, Speech and Signal Processing (ICASSP)},
  pages={6842--6846},
  year={2022},
  organization={IEEE}
}

@inproceedings{zou2024wave,
  title={Wave-mamba: Wavelet state space model for ultra-high-definition low-light image enhancement},
  author={Zou, W. and Gao, H. and Yang, W. and Liu, T.},
  booktitle={Proceedings of the 32nd ACM International Conference on Multimedia},
  pages={1534--1543},
  year={2024}
}

@inproceedings{wang2025mamba,
  title={Mamba-SEUNet: Mamba UNet for monaural speech enhancement},
  author={Wang, J. and Lin, Z. and Wang, T. and Ge, M. and Wang, L. and Dang, J.},
  booktitle={ICASSP 2025-2025 IEEE International Conference on Acoustics, Speech and Signal Processing (ICASSP)},
  pages={1--5},
  year={2025},
  organization={IEEE}
}

@inproceedings{jiang2025dual,
  title={Dual-path mamba: Short and long-term bidirectional selective structured state space models for speech separation},
  author={Jiang, X. and Han, C. and Mesgarani, N.},
  booktitle={ICASSP 2025-2025 IEEE International Conference on Acoustics, Speech and Signal Processing (ICASSP)},
  pages={1--5},
  year={2025},
  organization={IEEE}
}

@inproceedings{dang2024u,
  title={U-Mamba-Net: A highly efficient Mamba-based U-net style network for noisy and reverberant speech separation},
  author={Dang, S. and Matsumoto, T. and Takeuchi, Y. and Kudo, H.},
  booktitle={2024 Asia Pacific Signal and Information Processing Association Annual Summit and Conference (APSIPA ASC)},
  pages={1--5},
  year={2024},
  organization={IEEE}
}

@inproceedings{avenstrup2025sepmamba,
  title={SepMamba: State-space models for speaker separation using Mamba},
  author={Avenstrup, T. and Elek, B. and M{\'a}di, I. and Schin, A. and M{\o}rup, M. and Jensen, B. S. and Olsen, K.},
  booktitle={ICASSP 2025-2025 IEEE International Conference on Acoustics, Speech and Signal Processing (ICASSP)},
  pages={1--5},
  year={2025},
  organization={IEEE}
}

@inproceedings{wang2023tf,
  title={TF-GridNet: Making time-frequency domain models great again for monaural speaker separation},
  author={Wang, Z. and Cornell, S. and Choi, S. and Lee, Y. and Kim, B. and Watanabe, S.},
  booktitle={ICASSP 2023-2023 IEEE international conference on acoustics, speech and signal processing (ICASSP)},
  pages={1--5},
  year={2023},
  organization={IEEE}
}

@inproceedings{hershey2016deep,
  title={Deep clustering: Discriminative embeddings for segmentation and separation},
  author={Hershey, J. and Chen, Z. and Le Roux, J. and Watanabe, S.},
  booktitle={2016 IEEE international conference on acoustics, speech and signal processing (ICASSP)},
  pages={31--35},
  year={2016},
  organization={IEEE}
}

@inproceedings{le2019sdr,
  title={SDR--half-baked or well done?},
  author={Le Roux, J. and Wisdom, S. and Erdogan, H. and Hershey, J. R},
  booktitle={ICASSP 2019-2019 IEEE International Conference on Acoustics, Speech and Signal Processing (ICASSP)},
  pages={626--630},
  year={2019},
  organization={IEEE}
}

@article{wang2018supervised,
  title={Supervised speech separation based on deep learning: An overview},
  author={Wang, D.L. and Chen, J.},
  journal={IEEE/ACM transactions on audio, speech, and language processing},
  volume={26},
  number={10},
  pages={1702--1726},
  year={2018},
}

@article{lecun1989backpropagation,
  title={Backpropagation applied to handwritten zip code recognition},
  author={LeCun, Y. and Boser, B. and Denker, J. S and Henderson, D. and Howard, R. E and Hubbard, W. and Jackel, L. D},
  journal={Neural computation},
  volume={1},
  number={4},
  pages={541--551},
  year={1989},
}

@article{hochreiter1997long,
  title={Long short-term memory},
  author={Hochreiter, S. and Schmidhuber, J.},
  journal={Neural computation},
  volume={9},
  number={8},
  pages={1735--1780},
  year={1997},
}

@article{vaswani2017attention,
  title={Attention is all you need},
  author={Vaswani, A. and Shazeer, N. and Parmar, N. and Uszkoreit, J. and Jones, L. and Gomez, A. N and Kaiser, {\L}. and Polosukhin, I.},
  journal={Advances in neural information processing systems},
  volume={30},
  year={2017}
}

@article{luo2019conv,
  title={Conv-tasnet: Surpassing ideal time--frequency magnitude masking for speech separation},
  author={Luo, Y. and Mesgarani, N.},
  journal={IEEE/ACM transactions on audio, speech, and language processing},
  volume={27},
  number={8},
  pages={1256--1266},
  year={2019},
}

@article{li2022efficient,
  title={An efficient encoder-decoder architecture with top-down attention for speech separation},
  author={Li, K. and Yang, R. and Hu, X.},
  journal={arXiv preprint arXiv:2209.15200},
  year={2022}
}

@article{hu2021speech,
  title={Speech separation using an asynchronous fully recurrent convolutional neural network},
  author={Hu, X. and Li, K. and Zhang, W. and Luo, Y. and Lemercier, J. and Gerkmann, T.},
  journal={Advances in Neural Information Processing Systems},
  volume={34},
  pages={22509--22522},
  year={2021}
}

@article{li2022use,
  title={On the use of deep mask estimation module for neural source separation systems},
  author={Li, K. and Hu, X. and Luo, Y.},
  journal={arXiv preprint arXiv:2206.07347},
  year={2022}
}

@article{chen2020dual,
  title={Dual-path transformer network: Direct context-aware modeling for end-to-end monaural speech separation},
  author={Chen, J. and Mao, Q. and Liu, D.},
  journal={arXiv preprint arXiv:2007.13975},
  year={2020}
}

@article{kalman1960new,
  title={A new approach to linear filtering and prediction problems},
  author={Kalman, Rudolph Emil},
  year={1960}
}

@article{gu2021combining,
  title={Combining recurrent, convolutional, and continuous-time models with linear state space layers},
  author={Gu, A. and Johnson, I. and Goel, K. and Saab, K. and Dao, T. and Rudra, A. and R{\'e}, C.},
  journal={Advances in neural information processing systems},
  volume={34},
  pages={572--585},
  year={2021}
}

@article{gu2021efficiently,
  title={Efficiently modeling long sequences with structured state spaces},
  author={Gu, A. and Goel, K. and R{\'e}, C.},
  journal={arXiv preprint arXiv:2111.00396},
  year={2021}
}

@article{gu2022parameterization,
  title={On the parameterization and initialization of diagonal state space models},
  author={Gu, A. and Goel, K. and Gupta, A. and R{\'e}, C.},
  journal={Advances in Neural Information Processing Systems},
  volume={35},
  pages={35971--35983},
  year={2022}
}

@article{gu2023mamba,
  title={Mamba: Linear-time sequence modeling with selective state spaces},
  author={Gu, A. and Dao, T.},
  journal={arXiv preprint arXiv:2312.00752},
  year={2023}
}

@article{pioro2024moe,
  title={Moe-mamba: Efficient selective state space models with mixture of experts},
  author={Pi{\'o}ro, M. and Ciebiera, K. and Kr{\'o}l, K. and Ludziejewski, J. and Krutul, M. and Krajewski, J. and Antoniak, S. and Mi{\l}o{\'s}, P. and Cygan, M. and Jaszczur, S.},
  journal={arXiv preprint arXiv:2401.04081},
  year={2024}
}

@article{yang2024clinicalmamba,
  title={Clinicalmamba: A generative clinical language model on longitudinal clinical notes},
  author={Yang, Z. and Mitra, A. and Kwon, S. and Yu, H.},
  journal={arXiv preprint arXiv:2403.05795},
  year={2024}
}

@article{liu2024vmamba,
  title={Vmamba: Visual state space model},
  author={Liu, Y. and Tian, Y. and Zhao, Y. and Yu, H. and Xie, L. and Wang, Y. and Ye, Q. and Jiao, J. and Liu, Y.},
  journal={Advances in neural information processing systems},
  volume={37},
  pages={103031--103063},
  year={2024}
}

@article{shi2025vmambair,
  title={Vmambair: Visual state space model for image restoration},
  author={Shi, Y. and Xia, B. and Jin, X. and Wang, X. and Zhao, T. and Xia, X. and Xiao, X. and Yang, W.},
  journal={IEEE Transactions on Circuits and Systems for Video Technology},
  year={2025},
  publisher={IEEE}
}

@article{miyazaki2024exploring,
  title={Exploring the capability of mamba in speech applications},
  author={Miyazaki, K. and Masuyama, Y. and Murata, M.},
  journal={arXiv preprint arXiv:2406.16808},
  year={2024}
}

@article{chen2023neural,
  title={A neural state-space model approach to efficient speech separation},
  author={Chen, C. and Yang, C. and Li, K. and Hu, Y. and Ku, P. and Chng, E.},
  journal={arXiv preprint arXiv:2305.16932},
  year={2023}
}

@article{li2024spmamba,
  title={Spmamba: State-space model is all you need in speech separation},
  author={Li, K. and Chen, G. and Yang, R. and Hu, X.},
  journal={arXiv preprint arXiv:2404.02063},
  year={2024}
}

@article{wichern2019wham,
  title={Wham!: Extending speech separation to noisy environments},
  author={Wichern, G. and Antognini, J. and Flynn, M. and Zhu, L. R. and McQuinn, E. and Crow, D. and Manilow, E. and Roux, J. L.},
  journal={arXiv preprint arXiv:1907.01160},
  year={2019}
}

@article{cosentino2020librimix,
  title={Librimix: An open-source dataset for generalizable speech separation},
  author={Cosentino, J. and Pariente, M. and Cornell, S. and Deleforge, A. and Vincent, E.},
  journal={arXiv preprint arXiv:2005.11262},
  year={2020}
}
\end{footnotesize}


\end{document}